# GRAVITATIONAL CONUNDRUM: CONFUSING CLOCK-RATE MEASUREMENTS ON THE 'FIRST FLEET' FROM ENGLAND TO AUSTRALIA

**Richard de Grijs**
*School of Mathematical and Physical Sciences, Macquarie University,
Balaclava Road, Sydney, NSW 2109, Australia*
Email: richard.de-grijs@mq.edu.au

**Abstract:** Voyages of exploration often included astronomers among their crew to aid with maritime navigation. William Dawes, a British Marine who had been trained in practical astronomy, was assigned to the 'First Fleet', a convoy of eleven ships that left England in May 1787 bound for Botany Bay (Sydney, Australia). Dawes was also expected to take measurements of the local gravitational acceleration, *g*, at any port of call by measuring the daily rate by which his Shelton pendulum clock differed from that at Greenwich, its calibration location. Although Dawes and Nevil Maskelyne, Britain's fifth Astronomer Royal, had planned to obtain clock-rate measurements in the Canary Islands, San Sebastian (Rio de Janeiro) and Table Bay, Captain Arthur Phillip, Commander of the First Fleet, only allowed Dawes to disembark the clock in Rio de Janeiro. Therefore, we have just one set of clock-rate measurements from the voyage, in addition to land-based measurements obtained in New South Wales. If gravity was the dominant factor affecting the clock's changing rate, Dawes' measurement of −48.067 sec per (sidereal) day obtained in Rio de Janeiro implies a local gravitational acceleration, *g* = 9.7946 m sec$^{-2}$. On the other hand, if we adopt the modern value, *g* = 9.7878 m s$^{-2}$, the implied daily decay rate is almost exactly 30 sec greater than Dawes' clock-rate determination, a difference that is well in excess of the prevailing uncertainties. This suggests that the pendulum's regulator nut may have been offset by a full turn, thus implying that our assumptions regarding the pendulum length may have to be revisited.

## 1 GRAVITY EXPERIMENTS

The 'First Fleet' of 1787–1788 was a convoy of eleven ships that transported the first British convicts from England to Australia. As such, it marked the formal start of the British colonisation of the continent then known as New Holland. In 1786, Second Lieutenant William Dawes (1762–1836) of the British Marine Corps had been assigned to the First Fleet, in part as the expedition's dedicated astronomer (de Grijs and Jacob, 2021a,b).

In addition to Dawes' astronomical duties and his daily meteorological observations, we learn from his correspondence [1] with Nevil Maskelyne (1732–1811), Britain's fifth Astronomer Royal, that he also intended to take gravity measurements (e.g., Forbes, 1975: 172; Morrison and Barko, 2009; Bosloper, 2010): "I shall also pay particular Attention to the Experiment of Gravity, … The Rates of the Clock … shall be properly attended to …" (Dawes, 1787a). Dawes' Shelton regulator clock was clearly meant to provide gravity measurements, as we learn from his letter to Maskelyne of 1 October 1788, sent from their final destination in New South Wales:

> I fix'd very firmly the Clock in a niche of a solid Rock, on Sat$^y$. the 6$^{th}$ of Sept$^r$. and by Altitudes taken at two or three days interval found it to be losing at the rate of about 36"00 on sidereal time in one sidereal day but this Rate seems to be increasing … for in these last 11 Days the Clock has lost after the Rate of 37"25 on sidereal time in one sidereal day … (Dawes, 1788).

State-of-the-art gravity experiments in the late eighteenth century consisted of an extended series of daily clock-rate measurements of a carefully calibrated pendulum clock, compared with local time measurements based on astronomical observations. Dawes' measurements formed part of a comprehensive British research programme undertaken in the 1760s to 1780s of gravity measurements around the world. These included measurements obtained on H.M. Bark *Endeavour* by Charles Green (1734–1771) at Botany Bay during James Cook's (1728–1779) first voyage to the Pacific and Australia (1768–1771), and by William Bayly (1737–1810) and William Wales (1734?–1798) on H.M.S. *Resolution* in New Zealand, in 1773, during Cook's second voyage to the Pacific (1772–1775) (Bosloper, 2017).

At the time of the First Fleet's voyage, it had already been known for more than a century that pendulum swing ***periods*** vary with geographic latitude. As a case in point, in his *Philosophiæ Naturalis Principia Mathematica* (1687), Isaac Newton (1642–1726/7) commented that the French astronomer Jean Richer's (ca. 1630–1696) pendulum clock—which kept perfect time in Paris (latitude 48.86°N)—went slow in Cayenne (French Guyana; latitude 04.94°N):

> Now several astronomers, sent into remote countries to make astronomical observations, have found that pendulum clocks do accordingly move slower near the Equator than in our climates [at moderate northern latitudes]. And, first of all, in the year 1672, Mr. Richer took notice of it on the island of Cayenne; for when, in the month of August, he was observing the transits of the fixed stars over the meridian, he found his clock to go slower than it ought in respect of the mean motion of the Sun at the rate of $2^m 28^s$ a day. (Newton, 1687; see Figure 1).

Dawes was rather concerned that, upon his arrival at Botany Bay, the gradually accumulating errors in the Shelton pendulum clock's rate—possibly resulting from small deviations from isochronism[2]—would preclude him from performing the gravity experiments that had been proposed by Maskelyne:

> I wish to be informed how long a Time after our Arrival at Botany Bay, you think will be sufficient for making the Experiment of Gravity by the going of the Clock? it may reasonably be supposed to keep a very different Rate from that which it had at Greenwich, and I presume were it to be allowed to continue that Rate a long Time; the allowing for the Error of the Clock when that Error becomes very great will be rather inconvenient. (Dawes, 1787b: ff. 255r–v).

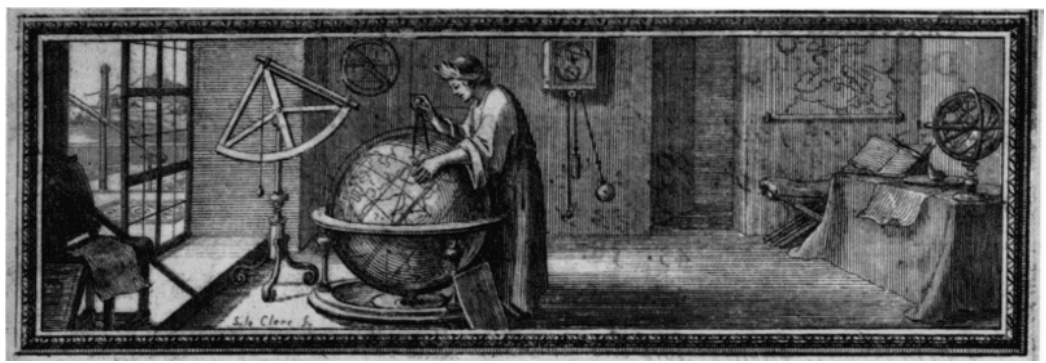

**Figure 1.** Drawing from Richer's *Observations astronomiques et physiques faites en l'isle de Caïenne* (Paris, 1679). Most of the astronomical instruments used by Richer are shown, including one of Isaac Thuret's (1630–1706) pendulum clocks. (Artist: Sebastian Le Clerc; Creative Commons Attribution-ShareAlike License.)

## 2 CLOCK REQUIREMENTS

For gravity calculations using pendulum clocks, the length of the temperature-compensated pendulum rod had to be precisely set and calibrated, and it also had to be reproducible anywhere, to within a tolerance of 10 µm. One also needed to measure the pendulum's half-amplitude—known as the 'arc from the vertical' or the 'arc of vibration'—so as to apply an amplitude-dependent correction to the clock-rate measurements. Maskelyne was well aware of these requirements, and he also knew how to calculate the amplitude-dependent correction; he made sure that all of his observers determined the 'arc from the vertical' of their pendulum clocks during their voyages, for a predetermined pendulum length that was set and calibrated at Greenwich.

As we saw already, Dawes' instruments included a Shelton astronomical regulator clock (see Figure 2), a highly precise, tall 'grandfather clock' equipped with a temperature-compensated grid-iron pendulum (see Figure 3) of precisely known length. Such regulator clocks were equipped with a so-called 'seconds' pendulum, which "… escaped dead seconds in the manner of the late Mr. [George] Graham" (Wales and Bayly, 1777) and completed a full swing period in 2 sec. They were calibrated to swing exactly 86,400 times during a mean solar day (24 hours, equivalent to 86,400 sec), which could be achieved by adjusting a regulator nut, that is, a fine-motion screw, at the bottom of the pendulum bob (see, e.g., the inset in Figure 2). These regulator nuts generally worked such that a full turn altered the clock rate by close to half a minute per day (Bosloper, 2010).

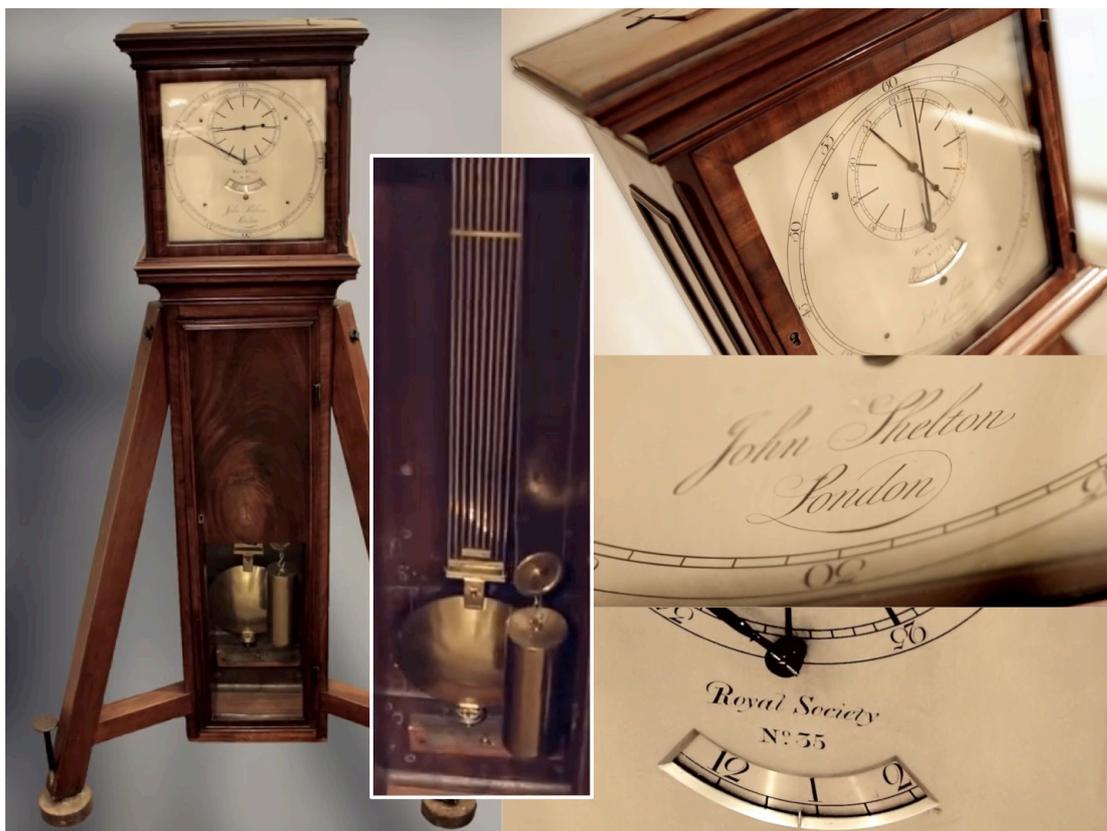

**Figure 2.** Shelton No. 35 regulator pendulum clock. (© The Royal Society. Screen captures from *Objectivity* video 57: https://www.youtube.com/watch?v=9fcWWBksfng. Reproduced for academic purposes under *YouTube*'s 'Fair Use' policy.)

Such astronomical regulator clocks were isochronal at their calibration location, usually Greenwich (at least initially). In 1733, James Bradley (1693–1762), Savillian Professor of Astronomy at Oxford University, had advised the Royal Society of London (Bradley, 1733) that the standard isochronal pendulum length was assumed to be 39.126 'English inches' (99.380 cm) (Bosloper, 2017: 248), a value attributed to the English clockmaker George Graham (1673/5–1751) in 1722. In 1817–1818, the British physicist and army captain Henry Kater (1777–1835) determined that the mean length of his 'reversible' solar seconds pendulum[3], in London, at sea level, in ambient conditions at a temperature of 62 °F (17 °C), swinging in a vacuum, was 39.13858 inches (99.412 cm) from knife edge to knife edge on his device (Kater, 1818). At the time of Dawes' observations during his voyage on the First Fleet and, subsequently, at Sydney Cove in New South Wales, Graham's value was still in use, however; it was adjusted to 39.128 inches (99.385 cm) in 1790 (*Ibid*.). We learn from a letter from Maskelyne to Henry Cavendish (1731–1810) in 1761 that the Astronomer Royal was fully aware of these developments (Maskelyne, 1761). Isochronal clocks marked precise 'dead' seconds at a given location, with the timing calibration provided by either stellar transit observations (for sidereal time) or local noon determinations (for mean solar time).

The length of the pendulum could, theoretically, be set precisely to a 'reduced length' of $g/\pi^2$ metres—where $g$ is the gravitational acceleration at one's reference location, e.g., Greenwich—simply by adjusting the regulator nut **when at that reference location**. The value of the gravitational acceleration ('absolute gravity'[4]) could then be determined, in principle, by multiplying the isochronal pendulum length by $\pi^2$, *modulo* a non-linearity correction.

In 1761, before he set off on his voyage to St. Helena to observe a Venus transit, Maskelyne had compiled a list of specifications for a pendulum clock to act as a frequency counter for gravity measurements (Higgit, 2014). Most importantly, it had to be equipped with a clearly visible and audible seconds hand. The requirement for the clock to tick once a second constrained the pendulum's length. In addition, the clock had to maintain power, that is, it could not suffer a loss of time when it was rewound, which was done approximately once a month.

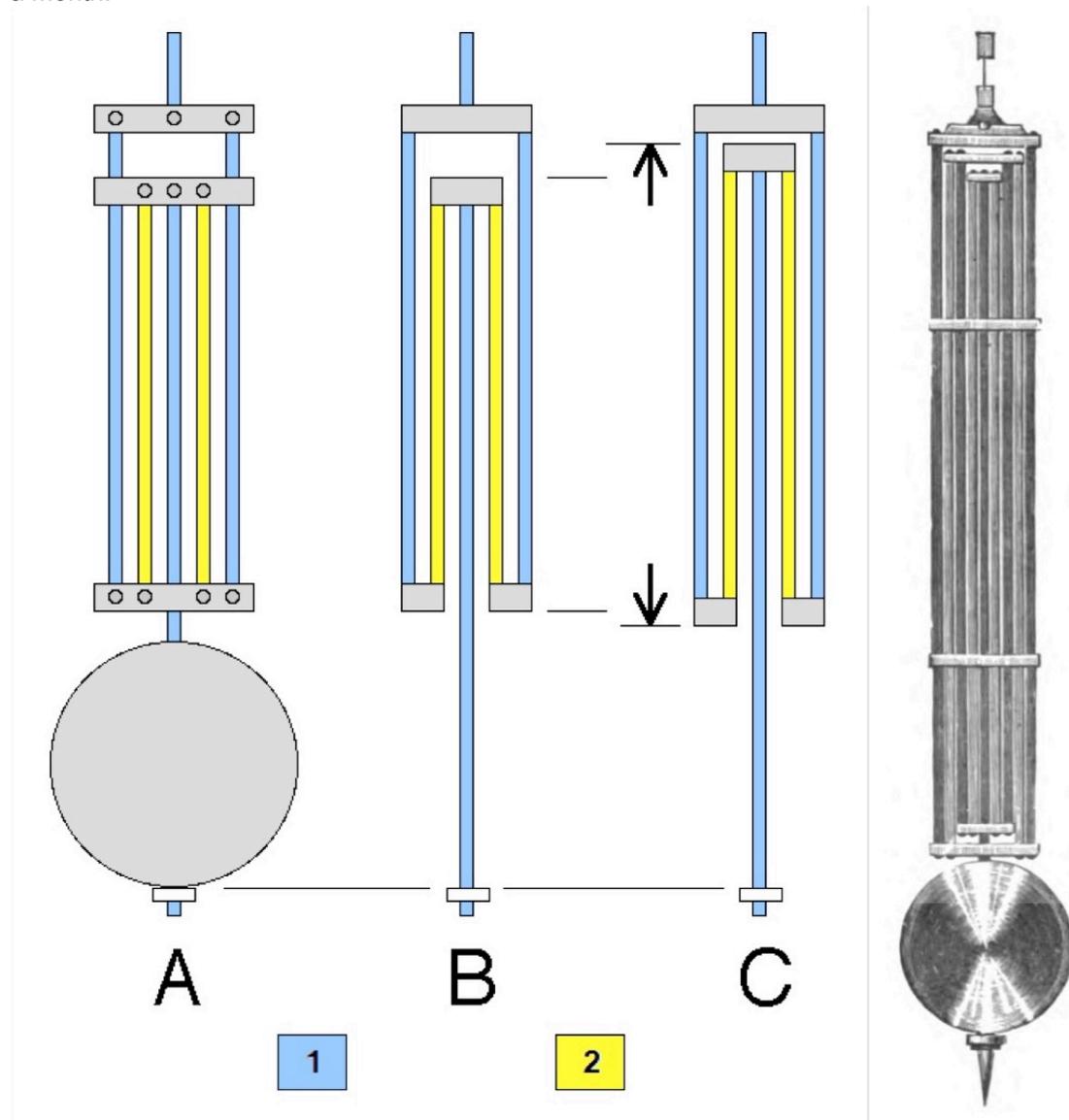

**Figure 3.** Operation of a temperature-compensated grid-iron pendulum. The pendulum uses rods of a high thermal-expansion metal, zinc (yellow) to compensate for the expansion of rods of a low thermal-expansion metal, iron (blue). A: General appearance; B: Schematic at normal temperature; C: Schematic at higher temperature (Creator: Leonard G., via Wikimedia Commons; public domain). The example pendulum on the right dates from the 1880s. Note the regulator nut at the bottom of the bob. (Brockhaus, 1890: Plate 124, Fig. 7, via Wikimedia Commons; public domain).

Finally, the pendulum had to be temperature-compensated. This was achieved by ensuring that the entire pendulum was composed of a bi-metal (see the grid-iron pendulum referenced above), an idea that had initially been proposed by Graham. It was famously applied by John Harrison (1693–1776) in the fourth incarnation of his 'longitude clock', H4, for which he was eventually awarded approximately half of the British Longitude Prize of 1714. The ratio of the metals' expansion coefficients in relation to their lengths had to be carefully balanced so as to provide full temperature compensation. In other words, expansion of one component of the bi-metal was meant to balance expansion of the other, thus keeping the pendulum length unchanged.

## 3 DAWES' MEASUREMENTS

While Maskelyne and Dawes were planning the scientific aspects of the First Fleet's voyage, they envisioned taking pendulum clock-rate measurements at all ports of call. Upon departure, Captain Arthur Phillip, Commander of the First Fleet and first captain of its flagship, H.M.S. *Sirius*, planned to call into the Santa Cruz roads at Tenerife, Canary Islands, as well as at St. Jago (Cabo Verde), San Sebastian (Rio de Janeiro) and Table Bay. The First Fleet did indeed call into Santa Cruz, San Sebastian and Table Bay, but a visit to St. Jago was abandoned at the last moment because of a sudden change in the prevailing winds that would have made entering the local harbour too dangerous.

Unfortunately, since Captain Phillip would not allow Dawes to disembark the Shelton regulator at either Tenerife or Table Bay, the Marine managed to obtain the pendulum clock's decay rate only in the harbour of Rio de Janeiro. In his letter of 3 September 1787, he went to great lengths to explain to Maskelyne his operational practice in San Sebastian while also pointing out a number of mishaps that had affected his measurements:

> The Reason why the mean Rate of the Clock is deduced [only] from the 16 to the 27.$^{th}$ [August 1787] is, that about 12 o'Clock at Night on the 17.$^{th}$ the Shoar [shore] mark'd in the Plan with a Stroke across it, was displaced, which I did not discover 'till the next Morning when I immediately replaced it, as nearly as I could judge in the same Positions that it had before; notwithstanding which I thought the mentioning of it indispensable, as also the not allowing the last Day's Rate which seems to have been affected by the Accident. I conclude that the small apparent Irregularity in the Rate of the Clock to be <u>merely</u> apparent, or very nearly so, and to arise principally if not entirely from the unavoidable Errors of Observation: I have therefore taken the mean Rate of the Clock – 48",067 in 24 sidereal, or 48",20 in 24$^h$ Mean Time for the Rate of the Clock each Day in the Comparisons of the Time Keeper with it. (Dawes, 1787c; original emphasis).

Dawes also provided information about his clock's regulator nut: "… the screw is at 15 on the nut …" (Dawes, 1788). However, Dawes only realised that the nut was set at that specific position following some confusion in Rio de Janeiro, potentially leading to erroneous results:

> Your Letter to me in which you mention the going of the Clock &c. says that the Index stood at 17 on the Nut;- when I opened the Clock Case, I found it exactly at 15 on which I immediately examined the line drawn on the Bar & found it but just appearing above the Bob, as if it (the Bob) had not been in the least lower'd; the measurement between the Bob & the upper End of the Brass Bar was accurately (as to Sense) the same as mentioned in your Letter and as I could in no wise account for the Nut having moved exactly 2 Divisions I thought there might possibly have been a mistake in taking down the Number on the Nut; I therefore set the Clock going with the Index exactly at 15 and on the next Page I have given the Results of each Day's Equal Altitudes. … On further consideration I do not think any Mistake could well have happen'd at Greenwich in noting what the Index stood at on the Nut & am rather fearfull [*sic*] that I ought to have screwed it to 17 before I set the Clock going; but in order to remedy as much as possible any ill Consequence that may have arisen from the not doing it, I intend the first Opportunity to try the Clock a Number of Days with the Index at 15 and afterwards to give it as fair a Trial as possible with the Index at 17. (Dawes, 1787c: ff. 269v, 272v).

Despite Dawes' valiant efforts, where he tried to determine whether the setting of the regular nut was meant to be 15 or 17, in practice this must have been nigh impossible. After all, the two nut positions would move the scratch mark by no more than 4 hundredths of a millimetre.[5] One would need to have access to properly calibrated daily decay rates at Greenwich, for both nut settings, to reach firm conclusions as to the original nut setting.

Eventually, Dawes' most accurate clock-rate observations were obtained once they had safely gained solid ground again at their final destination in New South Wales. As we saw already, at the time Dawes wrote to Maskelyne that he had observed that the Shelton regulator suffered a sustained loss of 37.25 sec per day over a period of 11 days—as it had done in Rio de Janeiro, he added (Dawes, 1788).

This information was all that Maskelyne needed to determine the gravity at Sydney Cove (present-day Circular Quay, adjacent to Sydney's central business district). Maskelyne knew which Shelton regulator Dawes had brought along with him on the voyage, as well as the precise length of its pendulum. It appears that he had set this length to 39.111 inches (99.342 cm), appropriate for a geographic latitude of 45°. In a letter dated 16 April 1790, from Sydney Cove, Dawes recorded that "… its arc of vibration is 1°30'+ and continues constantly the same …" (Dawes, 1790). This is the only recorded measurement of his pendulum clock's (half) amplitude obtained during the First Fleet's voyage. For isochronous operation, it applies to the entire voyage, however, as the half-amplitude is predominantly determined by a clock's escapement mechanism.

Modern analysis of Dawes' gravity measurements at Sydney Cove suggests that he indeed used a pendulum length of 39.111 inches, as opposed to the characteristic Greenwich length of 39.126 inches (Bosloper, 2010), corresponding to a difference of a full turn of the regulator nut (and a difference of 30 sec a day). By analogy with the gravity measurements undertaken by William Wales on Cook's second voyage using the same Shelton regulator, it has been suggested that Dawes made a similar 'mistake' as Wales (*Ibid*.). Wales recorded that the pendulum length

> ... was always altered, in order to its being packed up, yet on setting up again, it was constantly brought back to its proper length, by means of a scratch on the rod, and the numbers on the nut. (Wales and Bayly, 1777: 131).

However, in the introductory remarks to his observations from Cook's second voyage, he concedes,

> On reconsidering the circumstances of the clock's different rates of going at the Cape of Good Hope in November 1772 and April 1775, I am rather inclined to alter my opinion (see page 131) [previous citation] and to conclude that I made a mistake in setting the pendulum to its proper length, either when here in November 1772 or at Dusky Bay [now Dusky Sound] in New Zealand, after which time it was never altered; basically as the difference corresponds nearly to that which would arise from a whole revolution of the nut which supports the ball of the pendulum, namely 28" or 29", increased by the same quantity that the clock had gone faster on being set up a second time both at Point Venus [in Tahiti] and Queen Charlotte's Sound [New Zealand]. (Wales and Bayly, 1777: Introduction).

Dawes provided regular updates on the clock's rate of change during part of September 1788, after their arrival in New South Wales: on 1 October 1788 he advised Maskelyne that he "… had been so much employed in matters quite foreign to Astronomy that [he had] just had time to settle the going of the clock, …" (Dawes, 1788). In the same letter, he reported that the clock's rate was decreasing by 36 sec of sidereal time per day. He continued his measurements for most of his term in the colony, but his records were unfortunately misplaced after their transfer[6] to William Wales back in England.

The surviving measurements from September 1788, combined with a pendulum length ($L$) of 39.111 inches and local noon determined by 'equal altitudes' of the Sun, a standard approach to time measurement, result in a gravity at Sydney Cove of $g$ = 9.79705 m sec$^{-2}$, with an associated uncertainty of 25 parts per million (ppm) (Bosloper, 2010). The 1788 value resulting from Dawes' measurements is $g$ = 9.79621 m sec$^{-2}$ (Morrison and Barko, 2009: 34, Note 129). Combining this with the pendulum's arc of vibration of 1°30' leads to $g$ = 9.79705 m sec$^{-2}$, which is within 13 ppm of the modern value (*Ibid*.: Note 130), $g$ = 9.796720 m sec$^{-2}$ (Bell et al., 1973). Alternatively, based on the same pendulum length but adopting the change in the clock rate of –37.25 sec (slow with respect to the daily sidereal time) leads to a pendulum swing period of $T$ = 2.00086264 sec instead of exactly 2 sec (see below). Combined with the simple pendulum equation, $g = 4\pi^2 L/T^2$, we obtain $g$ = 9.79621 m sec$^{-2}$ at Sydney Cove (Morrison and Barko, 2009: 34, note 129).

We can now apply the same approach to obtain an estimate of the gravitational acceleration at Rio de Janeiro as implied by Dawes' measurements of 1787. For a given daily decay rate, the corresponding pendulum swing period $T$ would become slower than the standard 2 sec by a factor of 86,400/(86,400 – decay rate), where the decay rate is expressed as a positive value. Following Bosloper (2010), the gravitational acceleration can be expressed as

$$g = 4\pi^2 \frac{L(1+\Delta)^2}{T^2},$$

where $L$ = 39.111 inches = 0.9934194 m, and

$$(1 + \Delta)^2 = \left(1 + \frac{1}{2^2}\sin^2\frac{\theta_M}{2} + \frac{1}{2^2}\frac{3^2}{4^2}\sin^4\frac{\theta_M}{2} + \cdots\right).$$

For the arc of vibration pertaining to Dawes' Shelton regulator, $\theta_M = 1.5°$, it follows that $(1 + \Delta)^2$ = 1.000086 (Bosloper, 2010).[7] Combined with his measured decay rate of –48.067 sec with respect to the daily sidereal time, and assuming that the latter rate is entirely driven by the Earth's local gravity, this yields $g$ = 9.7946 m sec$^{-2}$ for Rio de Janeiro, presumably with a similar uncertainty of 25 ppm as for Sydney Cove.

**4 IMPLICATIONS**

The modern value for the gravitational acceleration in Rio de Janeiro is $g$ = 9.7878 m sec$^{-2}$ (Kochsiek and Gläser, 1999: 522). The gravitational acceleration implied by Dawes' clock-rate measurements is 0.0068 m sec$^{-2}$ larger than the modern value. This difference is well in excess of the probable uncertainty estimated by Bosloper (2010), although the accuracy of our measurements does not allow us to assess the difference to the parts-per-million level. At face value, this suggests either that gravity may not have been the only force acting on the Shelton regulator in Dawes' care or that one of our assumptions could be incorrect. Incomplete temperature compensation or any number of minor mechanical effects could have contributed to a slightly increased deceleration of the pendulum bob,[8] including non-negligible friction at the pivot point, an imbalance of the escapement mechanism or mechanical energy loss in the gear train.

On the other hand, let us consider which decay rate would be expected for the modern value of $g$ = 9.7878 m sec$^{-2}$. For the known values of $g$, $L$ and $(1 + \Delta)^2$, the corresponding pendulum swing period would be $T$ = 2.00180753 sec, corresponding to a declining clock rate of 78.1 sec per day. Now recall that Dawes had measured a decay rate of –48.067 sec per day in Rio de Janeiro. The difference between Dawes' measurement and the expected decay rate for the swing period implied by the modern value for the gravitational acceleration is almost exactly 30 sec per day.

Although we can no longer verify the actual setting of the regulator nut used by Dawes, a difference of 30 sec a day corresponds to a full turn of the regulator nut (Bosloper, 2010). Could it be that the clock's pendulum length had not been altered with respect to the characteristic Greenwich length after all?

A full turn of the nut changes the pendulum length by 0.66 mm, and one index number higher or lower on the nut only changes the pendulum length by 0.023 mm. If we now consider these implications with our current 20/20 vision, eighteenth-century observers seem to have found it hard to judge whether they were a full turn off from the proper setting of the nut, as witnessed by the errors made by Wales, Bayly and Dawes. That is most likely the reason why they also made a scratch mark on the pendulum rod just above the bob. One such scratch mark might refer to the calibration setting at Greenwich, another to an isochronal setting in a particular location where astronomical observations are timed. However, it was even harder to see if a change of two index steps on the nut—corresponding to a pendulum length change of 0.046 mm—made it disagree with the scratch mark, as Dawes noticed in Rio de Janeiro.

**5 NOTES**

[1] This correspondence is contained in the British Board of Longitude papers, held at the Cambridge University Library: https://cudl.lib.cam.ac.uk/view/MS-RGO-00014-00048/509

[2] 'Isochronism' refers to a clock's property when a pendulum's period is, at least for small swing angles, approximately independent of its (half) 'amplitude', that is, the maximum extent or angle of the swing.

[3] A reversible or 'Kater's pendulum' is a physical pendulum with two adjustable knife edges, purpose-built for accurate determination of the local gravitational acceleration, $g$.

[4] Strictly speaking, this gravity measurement was with respect to that at Greenwich, which became the *de facto* standard after Cavendish's publication of $G$, the 'gravitational constant', in 1798 (Cavendish, 1798).

[5] Although we can only speculate at this time, it is possible that the ship's vibrations during the voyage, as well as shocks, however minor, during the embarkation and disembarkation of the Shelton regulator at Portsmouth and in Rio de Janeiro may have caused such a minor shift in the nut's position.

[6] Upon the termination of Dawes' appointment in New South Wales and following a resolution by the Board of Longitude, on 3 June 1795 Maskelyne sent five of Dawes' "… folio manuscripts [journals] of Astronomical Observations made at Port Jackson and in the Voyage thither …" to Wales (Morrison and Barko, 2009: 22), Secretary of the Board of Longitude, for reanalysis and completion. Wales died before he had had the chance to complete and publish his work. Dawes' journals have been lost since that time (e.g., Bosloper, 2010).

[7] Note that this equation strongly implies that small changes in the pendulum's amplitude, e.g., from $\theta_M = 1°30'$ to $\theta_M = 1°40'$ or $\theta_M = 1°45'$ (e.g., Maskelyne, 1761–1762: 441; Cope, 1950: 261), will have a negligible effect on the resulting gravitational acceleration.

[8] The pendulum and bob of the No. 35 Shelton regulator operate inside a closed wooden case (information obtained from my own inspection of the clock at The Royal Society in London), so that any effects associated with air flow or air resistance can be ruled out.

## 6 ACKNOWLEDGEMENTS

I gratefully acknowledge a number of useful suggestions by the anonymous reviewers that helped me improve the robustness of my narrative.